\documentclass[runningheads,a4paper]{llncs}

\usepackage[utf8]{inputenc}
\usepackage[USenglish]{babel}
\usepackage[babel=false,english=american]{csquotes}

\usepackage{amssymb}
\usepackage{graphicx}
\usepackage[hyphens]{url}
\usepackage{listings}
\usepackage{float}
\usepackage{comment}
\usepackage{subfigure}

\usepackage{amsmath}

\usepackage{color}
\usepackage{xcolor}

\usepackage{url}
\urldef{\mailsa}\path|{francesco.gadaleta, nick.nikiforakis, jantobias.muehlberg , wouter.joosen}@cs.kuleuven.be|

\newcommand{\HRK}{\emph{Hello rootKitty}}

\usepackage{todonotes}

\begin{document}

\mainmatter
\title{HyperForce: Hypervisor-enForced Execution of Security-Critical Code}
\titlerunning{HyperForce: Hypervisor-enForced execution of security-critical code}

\author{Francesco Gadaleta, Nick Nikiforakis, Jan Tobias M{\"u}hlberg \\
        and Wouter Joosen}
\authorrunning{F. Gadaleta et al.}

\institute{IBBT-DistriNet, 
KU Leuven, Celestijnenlaan 200A B-3001, Leuven, Belgium\\
\mailsa\\
}

\toctitle{Lecture Notes in Computer Science}
\tocauthor{Authors' Instructions}
\maketitle

\begin{abstract}
The sustained popularity of the cloud and cloud-related services accelerate
the evolution of virtualization-enabling technologies. Modern off-the-shelf
computers are already equipped with specialized hardware that enables a
hypervisor to manage the simultaneous execution of multiple operating
systems. Researchers have proposed security mechanisms that operate within
such a hypervisor to protect the \textit{virtualized} operating systems from
attacks. These mechanisms improve in security over previous techniques
since the defense system is no longer part of an operating system's attack
surface. However, due to constant transitions between the hypervisor and
the operating systems, these countermeasures typically incur a significant
performance overhead.

In this paper we present HyperForce, a framework which allows the
deployment of security-critical code in a way that significantly
outperforms previous \textit{in-hypervisor} systems while maintaining
similar guarantees with respect to security and integrity.  HyperForce is a
hybrid system which combines the performance of an \textit{in-guest}
security mechanism with the security of in-hypervisor one.  We
evaluate our framework by using it to re-implement an invariance-based
rootkit detection system and show the performance benefits of a HyperForce-utilizing 
countermeasure.
\end{abstract}


\section{Introduction}


The ``cloud'' is probably the most used technological term of the last years. Its supporters present it as a complete change in the way that
companies operate that will help them scale on-demand without the hardware-shackles of the past. CPU-time, hard-disk space, bandwidth and complete
virtual infrastructure can be bought at a moment's notice. Backups of data are synced to the cloud and in some extreme cases, all of a user's data may reside
there (Chromium OS). Its opponents treat it as privacy nightmare that will
take away the user's control over their own data and place it in the hands
of corporations as well as a risk to the privacy, integrity and availability of user data~\cite{amazon_cloud_problem,exposingFHS2011}. Regardless however on one's view of the cloud, one of the  main technologies that 
makes the cloud-concept possible is virtualization.

Virtualization is the set of technologies that together allow for the existence of more than one running operating systems on-top of a single physical machine.
While initially all of the needed mechanisms for virtualization were created in software, the sustained popularity of virtualization, lead to their implementation in hardware, providing
the desired speed that was lacking in their software counterparts. Today both Intel\footnote{\url{http://www.intel.com/technology/virtualization/technology.htm}} and AMD\footnote{\url{http://sites.amd.com/us/business/it-solutions/virtualization}} support a set of instructions that are there with the sole purpose of facilitating 
virtualization. Apart from the use of virtualization as way to host different operating systems on one machine, virtualization can also be used to provide 
greater security guarantees for operating systems. Researchers have already proposed various system that use virtualization primitives that all fall in this category
~\cite{Criswell2007,Dewant2008,hellorootkitty,Gadaleta2009,Qubesos}. The chief difference between these systems that operate from within the virtualized system's hypervisor (\textit{in-hypervisor}) 
and protection systems that operate from within the operating system (\textit{in-guest}) is
that the latter are part of the system's attack surface. For instance, an
antivirus that operates from within the operating system that it supposedly
protects, i.e. in-guest, could
be deactivated or crippled by the attack itself. In-hypervisor
security systems, in contrast, can utilize the isolation guarantees of virtualization to make sure that they will
be active regardless of the state of system that they protect. Unfortunately these security benefits do not come for free. The constant transition from the virtualized 
operating system to the hypervisor that protects it (known as \texttt{VMExit}) and back (\texttt{VMEntry}), negatively affects the performance of the virtualized systems forcing one to choose between better security or better performance.

In this paper we present HyperForce, a framework that allows countermeasures for virtualized operating systems to be protected, with security and integrity comparable to the one provided by in-hypervisor systems but at the performance cost of in-guest systems. Our system follows a hybrid approach by maintaining the security-critical code within the guest but 
forcing its execution and protecting its instructions and data from the hypervisor. Using our framework, we re-implement Hello rootKitty~\cite{hellorootkitty}, an in-hypervisor
rootkit-detection system which uses the invariance of critical kernel objects as a way of identifying kernel compromises. We evaluate the implementation of Hello rootKitty using 
our framework and show that it significantly outperforms the original version while maintaining comparable security guarantees.

The rest of this paper is structured as follows. In Section~\ref{sec:design} we present our motivation for the HyperForce framework followed by its design 
details. In Section~\ref{evaluation} we evaluate the performance benefits of our framework by using it to re-implement the aforementioned rootkit-detection system. In Section~\ref{sec:related} we
discuss the related work and Section~\ref{sec:conclusion} concludes.


\section{Design}
\label{sec:design}
In this section, we describe the needs that motivated us to create HyperForce and we provide the design and implementation details of our system.

\subsection{Motivation}
Designing a countermeasure that protects virtualized operating systems is considered a challenge not only because of the difficulty to modify the target system (due to the lack of sources or licenses) but also because a virtualized system is already affected by consistent overhead, by design.
An important goal for any framework using virtualization as a security tool, is to guarantee the execution of critical code in the kernel-space of a virtualized operating system regardless of the state of the kernel, i.e. code that will run identically in both clean and compromised kernels.
By critical code we refer to code that, in general, monitors the state of the system and that it is desirable, mainly from a security point of view, to maintain its execution.
Examples of such code include the integrity check of sensitive kernel-level data structures that are usually abused by rootkits or the scanning of 
files and memory for known malware signatures. Given our assumption of a kernel-level attacker, it is also needed to ensure the integrity of the critical code to protect
it from malicious modifications which might compromise its efficacy or completely disable its operations.

A straightforward way of achieving this goal is to implement and execute security-critical code within the hypervisor \cite{hellorootkitty}. An alternative approach monitors the 
target system from a separate virtual machine. In fact, one of the most interesting features of virtualization technology is that it guarantees complete isolation between the hypervisor and any virtual machine running on top of it as well as isolation between multiple virtual machines running on top of the same physical machine.
Unfortunately, both approaches are known to be affected by consistent performance overhead, making it hard to consider such solutions for production systems.
The main goal of HyperForce is to keep a degree of security comparable to these completely isolated systems while significantly reducing their performance overhead.

\subsection{Core Idea}
The idea of HyperForce is to combine the best features of the \textit{in-guest} and \textit{in-hypervisor} defense systems into a hybrid solution which performs as an in-guest countermeasure
while providing security comparable to in-hypervisor countermeasures. We achieve this by deploying the functional part of the countermeasure within the guest operating system while maintaining its integrity and enforcing its execution with the assistance of the hypervisor. Since the functional part of the security-critical code, i.e. its instructions and data, is running within the
virtualized operating system, it also has native access to the resources of the virtualized operating system such as the memory, disk and API of the virtualized kernel. 
This provides a great performance benefit for code that needs to access many memory locations within the virtualized operating system since it alleviates the costly need of introspection 
that in-hypervisor systems require, i.e. the discovery of the corresponding physical memory pages of the virtual memory pages of the guest and their remapping  within the 
hypervisor or within another virtual machine. 


\subsubsection{Enforcement of Execution.}
Given an arbitrary piece of security-critical code, HyperForce needs to ensure its execution at regular time intervals. A complete reliance for its execution on the
virtualized operating system, could potentially allow a kernel-level attacker to intervene and inhibit the code's execution through the modification of the appropriate 
kernel-level data-structures. For instance, an attacker could locate the function pointer pointing to the security-critical code and overwrite it with a pointer towards 
their own code.

From a high-level view, HyperForce changes the execution flow of the guest kernel whenever the installed monitoring code has to be executed and restores the original 
execution flow upon code termination. The advances of virtualization technology allows one to implement this transition in a multitude of ways. Our decision was influenced by
our desire of minimizing the amount of instrumentation code in the hypervisor and of keeping performance overhead to a minimum.

\begin{figure} 
\begin{center}
\includegraphics[scale=0.5]{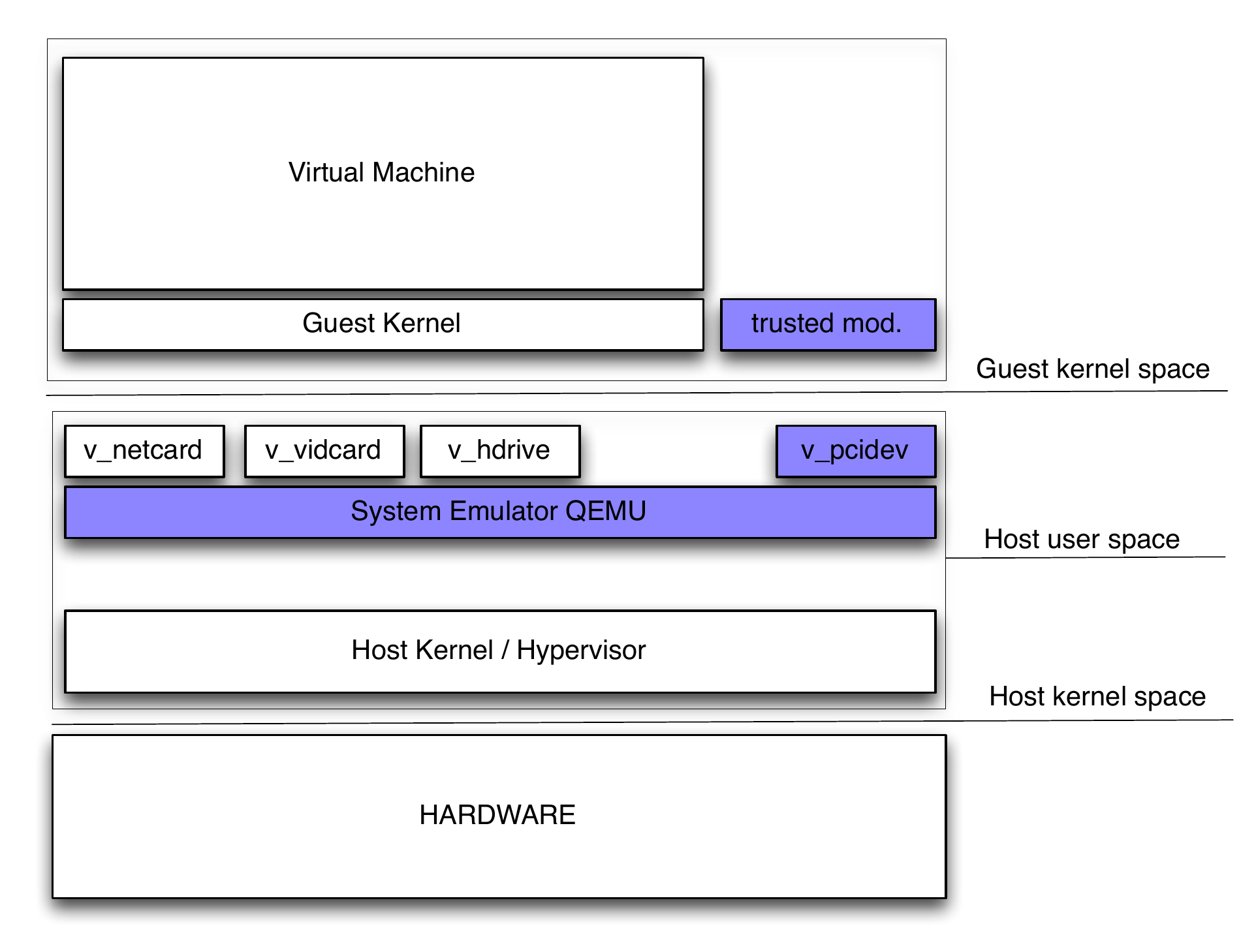}
\caption{Schema of HyperForce. Highlighted components indicate parts of the system that need instrumentation/modification}
\label{hyperforce_schema}
\end{center}
\end{figure}

In our framework, the security-critical code is encapsulated within a function that is loaded in the virtualized operating system in the form of Linux Kernel Module (LKM). This allows
the code to have native access to all of the VM's native resources.
HyperForce then uses the infrastructure of the virtualization platform, specifically the Virtual Machine Monitor (VMM), to create a virtual device. Virtual devices simulate real hardware
devices, such as sound-cards and video cards, and are supported by all modern Virtual Machine Monitors. Once this virtual device is created and loaded in the virtualized operating system,
HyperForce then registers the address of the security-critical code as an interrupt handler for the virtual device, as illustrated in Figure~\ref{hyperforce_schema}. 
The cooperation of the hypervisor and the trusted module, allows for the security-critical code to execute every time that
the virtual device generates an interrupt. 

Since the virtual device is fully controlled by the hypervisor, it is the hypervisor that decides when interrupts must be generated and
not the virtualized operating system. Due to this fact, the possibly compromised kernel of the VM, cannot anticipate when the security-critical code will be executed since the logic behind it
is hidden from it through the virtualization-guaranteed isolation between hypervisor and VM. This fact stops any attackers' efforts to evade detection by mimicking a non-compromised operating system just before the execution of the critical-code and restoring their malicious activities after it.


\subsubsection{Integrity of Code.}
In the previous paragraphs we described the loading of security-critical code within the hypervisor and the use of virtual devices to ensure the execution of that code.
Since the code is loaded in the VM as a LKM, it executes with the privileges of the virtualized kernel. While this is desired, it also opens up the code to attacks, e.g. modifications of
its code and data, from an attacker who is in control of the virtualized
kernel. Traditionally, the module could not be protected from the rest of the kernel since they both operate within 
the same protection ring, namely Ring 0. Due to virtualization however, the hypervisor has more power than the virtualized operating system's kernel (signified as Ring -1) and can
thus protect any resources from the virtualized kernel, including memory pages. HyperForce takes advantage of this fact, and write-protects the memory pages holding the instructions
and data of the security-critical code. In order to allow the code to make changes to its data, HyperForce can unlock the memory pages before it triggers an interrupt of its virtual device
and lock them back immediately after the code's execution. In order
to ensure that an attacker cannot avoid the execution of the interrupt handler containing the security-critical
code, HyperForce also write-protects the memory page holding the Interrupt Descriptor Table (IDT) of the protected VM. Lastly, HyperForce protects the Interrupt Descriptor Table Register (IDTR) that contains the address of the IDT, as a regular invariant critical kernel object.

\section{Evaluation} \label{evaluation}
We implemented HyperForce in KVM, an extension of the Linux kernel with hypervisor capabilities. KVM is formed by a system emulator, QEMU, that runs as regular process in user space and a kernel-space device driver that uses virtualization-enabled processors.
In order to show the improvements provided by our framework we chose to re-implement and measure a pure \textit{in-hypervisor} countermeasure, namely \HRK{}~\cite{hellorootkitty}.


\HRK{} is a lightweight invariance-enforcing framework that mitigates the problem of kernel-level rootkits. 
It represents a typical in-hypervisor monitoring system that checks the integrity of invariant guest-kernel objects from the hypervisor. A periodical mapping of guest-kernel memory into hypervisor space is followed by computation of its hash and checks against a set of precomputed values.
Such a countermeasure often deals with a high number of kernel objects and performance overhead can easily make the guest system unusable.  
To minimize the amount of time spent by additional code, only a subset of these objects is checked whenever control returns to hypervisor (VMExit). Thus a certain number of VMExit events is needed to check the entire list of protected objects. This relaxation will have a cost in terms of detection time needed to check the entire list of objects.

\begin{figure}
\begin{center}
\includegraphics[scale=0.5]{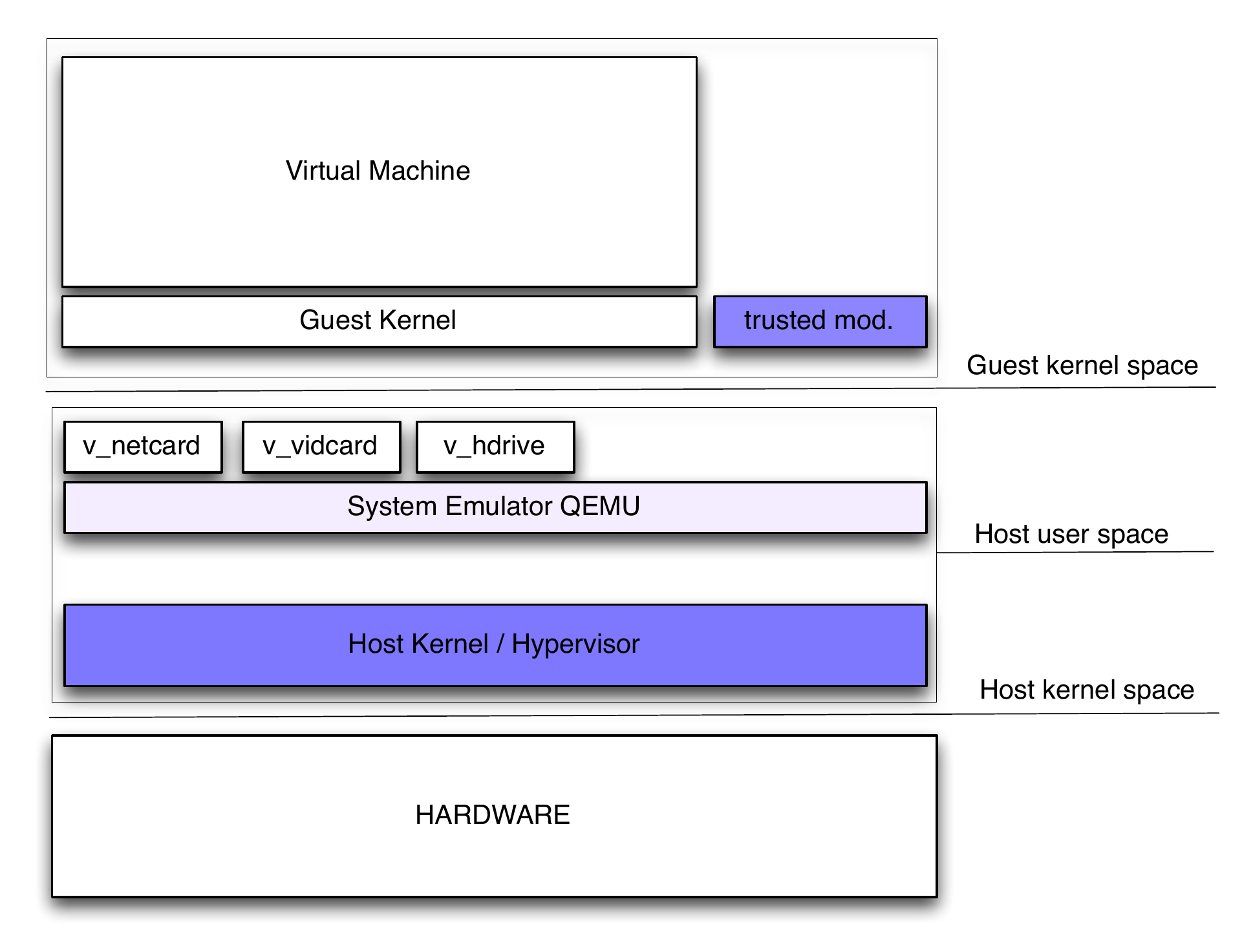}
\caption{Schema of \HRK{} implemented in Linux KVM. Highlighted components indicate parts of the system that need instrumentation/modification}
\label{hellorootkitty_schema}
\end{center}
\end{figure}

The original version of \HRK{} was implemented in BitVisor~\cite{1508311}, a tiny hypervisor designed for mediating I/O access from a single guest operating system. In order to be able to
fairly compare it with our implemented version using HyperForce, we also re-implemented the original \HRK{} in KVM.
A schema of \HRK{} implemented in KVM is provided in
Fig.~\ref{hellorootkitty_schema}. It can be observed that while the
HyperForce framework requires only the system emulator to be modified
(Fig.~\ref{hyperforce_schema}), the implementation of the in-hypervisor \HRK{} needs instrumentation code to be added to the host kernel. In both cases the trusted module needs to be added to the guest kernel.\\

We collected results of macro and micro-benchmarks from the guest and from the host machine and discuss them in Section \ref{macrobench} and Section \ref{microbench}.
In order to provide reliable results, all tests have been repeated 10 times and averaged. Experiments have been performed on Intel Core 2 Duo 2 Ghz processor with 4GB of RAM.


\begin{table}[t]
\centering
\subfigure[In-host measurements]{
\label{hostmacro}
\begin{tabular}{| l | l | l |}
\hline
     ~                 & iperf [Gb/s] & overhead \\
\hline
 &  & \\
HRK      & 6.36         &	-               \\
HF(HRK)		& 6.29         &	+1.1\%    \\
\hline
\end{tabular}
}
\subfigure[In-guest measurements]{
\label{guestmacro}
\begin{tabular}{| l | l | l |}
\hline
     ~                 & iperf [Gb/s] & bunzip [sec] \\
\hline
native KVM	& 5.97       &	32.04        \\
\hline
HRK      & 5.26   (+12\%)      &	33.73  (+5\%)       \\
HF(HRK)		& 5.71   (+4.3\%)      &	32.88  (+2.5\%)    \\
\hline
\end{tabular}
}
\caption{Macro benchmarks (\textit{in-host} OS and \textit{in-guest} OS) evaluating \HRK{} implemented with and without HyperForce}
\label{macrobenchmarks}
\end{table}

\subsection{Macro-benchmarks} \label{macrobench}
We run two macro benchmarks, \emph{iperf} that measures TCP and UDP bandwidth performance and \emph{bunzip} of a Linux kernel source code. The original in-hypervisor version of \HRK{} is denoted as ``HRK'' while the version using the HyperForce framework is denoted as ``HF(HRK)''.

While the in-hypervisor approach, due to the slower context switching, has
a slightly better throughput of network performance in the host machine
Table~\ref{macrobenchmarks}(a), benchmarks in the guest machine show a considerably better performance with HyperForce.   
\emph{iperf} and \emph{bunzip} have also been executed on a native KVM system and compared against the same system running in-hypervisor \HRK{} and then HF(HRK). 
The performance overhead of our approach is about half of the in-hypervisor
\HRK{}, as shown in Table~\ref{macrobenchmarks}(b). \HRK{} implemented
using HyperForce performs with 4.3\% overhead compared to a native KVM guest while in-hypervisor \HRK{} shows 12\% overhead.
The second column reports overhead of \textit{bunzip} measured in seconds. HF(HRK) outperforms the in-hypervisor \HRK{}, showing an overhead of only 2.5\% compared to the native KVM guest.

\subsection{Micro-benchmarks} \label{microbench}
Micro benchmarks show a more detailed picture of the two approaches. 
We use \emph{LMbench}~\cite{lmbench-paper}\footnote{We use version 3 of
LMbench as available at \url{http://lmbench.sourceforge.net/}.} to measure the overhead of operating system specific events such as context switch, memory mapping latency, page fault, signal handling and fork.
Within the host machine, HyperForce shows substantial improvement against the alternative \HRK{}. 
In Table~\ref{hostmicro} we report only the tests where this improvement is consistent. In all other tests the in-hypervisor \HRK{} and the \HRK{} using HyperForce show negligible performance overhead. 

The picture in the guest machine shows a similar trend in which HF(HRK) outperforms the original in-hypervisor \HRK{} in every test (Table~\ref{guestmicro}).

\begin{table}[htdp]
\caption{Overhead of \HRK{} using the HyperForce framework (HF(HRK)) is
measured against in-hypervisor \HRK{} (HRK) with LMbench micro-benchmarks within the host machine. Operations are measured in microseconds.}
\begin{center}
\begin{tabular}{| l | c | c | c | c | }
\hline
~ & ctx switch  & mmap lat  & page flt 	 & mem lat \\
\hline
HRK     & 2.020  &	6148	&	1.57		&114.7  \\
HF(HRK)		& 1.48    &	4950	&	1.46		& 101.7 \\
\hline
speedup         &+26\% & +19\% & +7\% & +11\% \\
\hline
\end{tabular}
\end{center}
\label{hostmicro}
\end{table}%

\begin{table}[htdp]
\caption{Overhead of HF(HRK) is measured against in-hypervisor \HRK{} with
LMbench micro-benchmarks within the guest machine. Operations are measured in microseconds.}
\begin{center}
\begin{tabular}{| l | c | c | c | c | c | c | c | c |  }
\hline
  ~            & null call  & null IO & open/close & sig inst   & sig handl & fork proc & exec proc & ctx switch \\ 
\hline
HRK & 0.30 & 0.32 & 2.32 & 0.74 & 5.37 & 1923 & 4087 & 5.58  \\
HF(HRK)    & 0.14 & 0.21 & 2.10 & 0.45 & 2.60 & 1788 & 3984 & 5.00  \\  
\hline
\hline
Speedup      &  +53\%   & +34\%    &  +10\%    & +39\%   & +51\%    & +8\%       & +2.5\%     &  +10\%   \\
\hline
\end{tabular}
\end{center}
\label{guestmicro}
\end{table}%

To interpret the results shown in Table~\ref{guestmicro}, one has to know
that \HRK{} performs integrity checks whenever the guest kernel writes to a
control register (\texttt{MOV\_CR*} event). When virtual addressing is
enabled, the upper 20 bits of control register 3 (\texttt{CR3}) become the
page directory base register, which is used to locate the page directory and
page table of the current process. Thus, on every context switch or system
call invocation, CR3 is modified. Trapping these events strategically
contributes to \HRK's short attack detection time. Yet, the integrity
checks performed by \HRK{} increase the latency of context switches and
system calls.

In contrast, our implementation of \HRK{} in HyperForce employs interrupt
events to trigger in-guest integrity checks. This eliminates overheads with
respect to switching execution context and address mapping between the
hypervisor and the guest OS, while the remaining computational overhead
affects guest operations more evenly. As can be seen in
Table~\ref{guestmicro}, the above changes imply significant speedups on
system call invocations (53\%) and context switches (10\%). Although
LMbench is often considered as insufficient for evaluating system
performance~\cite{lmbenchevil}, our example shows that the benchmark suite
can be used to neatly distinguish the actual speedup on system call
invocations (\enquote{null call}) from the impact on a particular system
call execution (e.g. \enquote{open/close}).

One may think that our approach to trigger security checks through interrupts in
HyperForce reduces the security of the protected system compared
to the original in-hypervisor \HRK{}: in the latter case an attacker increases their
chance of being detected with every system call raised. However for a total
of 15,000 protected kernel objects, the worst-case detection time reported
in~\cite{hellorootkitty} is 6 seconds. \HRK{} in HyperForce improves on
that by checking the same amount of kernel objects in 4 seconds. While \HRK{} 
relies on the activity of the system as a trigger that checks the integrity of
protected objects, \HRK{} in HyperForce performs the checking independently of 
system activity every 4 seconds. 

In summary, our implementation of \HRK{} in HyperForce significantly reduces
computational overhead while reducing the worst-case detection time for
potentially malicious manipulations of invariant kernel objects. Our results
indicate that the HyperForce framework could be used to re-implement other
in-hypervisor applications, enhancing their performance and maintaining their effectiveness.

\section{Related Work}
\label{sec:related}
In this section we review related work in the domain of kernel code
integrity assurance. For a discussion of literature related to rootkit
detection we refer the reader to \cite{hellorootkitty}. 
%

\subsubsection{Hardware-Based Execution Flow Integrity.} Means of guaranteeing
the integrity of a running operating system that employ dedicated hardware
devices to monitor the physical memory of a computer system have been
proposed in \cite{gibraltar} and \cite{copilot}. In order to perform
integrity checks, both systems make use of PCI hardware that directly
accesses the computer's memory at a negligible performance overhead. Yet,
the need for dedicated hardware may hinder widespread deployment of these
techniques.

\subsubsection{Hypervisor-Based Execution Flow Integrity.} A tiny hypervisor
that protects legacy OSs by ensuring that only validated code can be
executed in kernel mode, is SecVisor~\cite{SecVisor}. A similar system,
NICKLE~\cite{NICKLE}, shadows physical memory to store authenticated guest
code. At runtime, an instruction fetch is directed to access either the
normal system memory or the shadow area, depending on whether the
instruction is to be executed in user mode or kernel mode. An attempt to
execute unvalidated code can thus be detected and prevented.
Recently, attacks that do not inject malicious code but construct it from
existing fragments of the attacked program have been presented
\cite{goodbad,Hund2009,geometry}. These attacks effectively bypass
countermeasures such as SecVisor and NICKLE. 

Rootkits commonly modify a system's function pointers to ensure execution.
HookScout~\cite{hookscout} detects such rootkits. The tool employs as
system emulator to infer a policy for function pointer propagation in
kernel memory. A separate detection system is then used to detect
violations of this policy during normal operation of the OS. Since the
detection system runs on the target machine, it may be disabled by an
attack.
HookSafe~\cite{HookSafe} protects kernel hooks that are dynamically
allocated by relocating these kernel hooks to dedicated memory pages.
Regular page-level protection through the hardware's Memory Management Unit
is then used to protect the pages. Yet, the technique does not prevent
non-control data from being compromised.

HyperForce can be utilized to effectively protect the integrity of the
in-guest components of systems such as HookScout and HookSafe. Our
experimental results obtained from implementing \HRK~\cite{hellorootkitty}
in HyperForce show that our technique leverages the use of in-guest
protection mechanisms. That is, HyperForce substantially reduces the
performance overhead that would occur if the countermeasure would be
implemented in-hypervisor, while strong security guarantees are maintained.

\subsubsection{Security Agent Injection.} Closely related to HyperForce is
work by Lee et al.~\cite{lee:agent-injection} and Chiueh et
al.~\cite{chiueh:sade} on deploying agents by means of code injection from
a hypervisor. Both approaches are applicable to guest OSs that have not
been previously prepared by loading a special driver or similar.
In \cite{lee:agent-injection}, Lee et al. proposes to protect agent code
that is executing in a compromised guest OS kernel by the use of
cryptography and by injecting this code on demand from the hypervisor. As
there is no implementation and no experimental evaluation given, a
comparison with HyperForce is not feasible.
Similarly, work on SADE~\cite{chiueh:sade} by Chiueh et al. uses VMWare's
ESX server API to inject and execute code in a guest OS so as to disable
and remove a previously detected malware infection from that guest. In
difference to the HyperForce approach, the agent code in SADE is not
protected from malicious interference on the guest. Chiueh et al. argue
that on-demand injection leaves a relatively short time span for such
interference. SADE is used by a virtual appliance that implements
out-of-guest monitoring of VMs' memory, scanning for malware signatures.
The paper presents experimental data on the code injection process but does
not discuss the overhead implied by mapping memory pages between the
virtual appliance and the VMs. We expect in-guest memory inspection, as
implemented by our Hello rootKitty in HyperForce, to outperform SADE.

\section{Conclusion}\label{sec:conclusion}

The attractive properties offered by virtualization are a foundational
block for the whole ``cloud technology''. At the same time, virtualization
is already being used for purposes other than the deployment of multiple
operating systems as a way of increasing the security of a single
virtualized operating system. In this paper we briefly discuss the
differences between security mechanisms deployed within an operating system
(\textit{in-guest}) and the ones deployed within a hypervisor
(\textit{in-hypervisor}) and bring attention to the, seemingly exclusive,
choice between the performance benefits of the former versus the security
benefits of the latter. We tackle this choice by developing HyperForce, a
hybrid framework allowing security mechanisms to be developed in a way that
provides them with performance analogous to in-guest systems while
maintaining the security of in-hypervisor systems. Using
HyperForce, we re-implemented an in-hypervisor rootkit detection
system and show how the new version significantly outperforms the original
without compromising the security or integrity of the detection system.

We conclude that hybrid security systems that are built on top of
HyperForce can provide effective and efficient alternatives to mitigate the
overhead of techniques that exclusively operate in-hypervisor.
Interesting candidate applications for our framework are, e.g., malware
detection and removal software. For future work we envisage to extend
HyperForce with techniques to inject security agents into a guest operating
system so as to provide secure means of on-demand deployment of such
agents.

\paragraph{Acknowledgments.} This research is partially funded by the
Interuniversity Attraction Poles Programme Belgian State, Belgian Science
Policy, the Research Fund KU Leuven and the EU FP7 project NESSoS.


\bibliographystyle{plain}
\bibliography{hyperforce}

\end{document}